# Polylogarithmic representation of radiative and thermodynamic properties of thermal radiation in a given spectral range: I. Blackbody radiation

Anatoliy I Fisenko, Vladimir Lemberg

*ONCFEC Inc., 250 Lake Street, Suite 909, St. Catharines, Ontario L2R 5Z4, Canada*

E-mail: *afisenko@oncfec.com*

**Abstract** Using polylogarithm functions the exact analytical expressions for the radiative and thermodynamic properties of blackbody radiation, such as the Wien displacement law, Stefan-Boltzmann law, total energy density, number density of photons, Helmholtz free energy density, internal energy density, enthalpy density, entropy density, heat capacity at constant volume, and pressure in the finite range of frequencies are constructed. The obtained expressions allow us to tabulate these functions in various finite frequency bands at different temperatures for practical applications. As an example, the radiative and thermodynamic functions using experimental data for the monopole spectrum of the Cosmic Microwave Background (CMB) radiation measured by the *COBE* FIRAS instrument in the 60 - 600 GHz frequency interval at the temperature $T$ = 2.725 K are calculated. The expressions obtained for the radiative and thermodynamic functions can be easily presented in wavelength and wavenumber domains.

# 1 Introduction

In almost all areas of physics and astrophysics related to the measurement of spectral and integral characteristics of the emitted systems by using devices such as Fourier Transform InfraRed (FTIR) spectroscopy, Fourier Transform spectroscopy, Far InfraRed Absolute Spectrophotometry (FIRAS), ultraviolet-visible (UV-Vis) spectroscopy, high temperature radiance spectroscopy, dispersive spectrometer, fast spectro-pyrometer, infrared scanners, etc., a small region of the electromagnetic spectrum has a practical interest. These instruments have been developed to determine the spectral radiation intensity seen at short radiation windows. For examples: a) to measure human skin temperature using infrared scanners the wavelength between 2μm and 6μm is of practical interest [1]; b) the range of wavenumber from 2 to 20 cm$^{-1}$ for the measurement of the Cosmic Microwave Background (CMB) radiation [2] was used in the *COBE* FIRAS observations; c) in [3], an experimental study of the radiative properties of liquid and solid uranium and plutonium carbides in the finite wavelength range of 550 nm $\leqslant \lambda \leqslant$ 920 nm was the main practical interest; and many others.

Theoretical experiments using a computer require to obtain the numerical solution for the calculation of the total thermal radiation, number density of photons as well as the thermodynamic functions over a specified range of the spectrum.

It is well-known that in the study of radiative heat transfer the blackbody is used as an ideal theoretical standard with which the absorption of real bodies compared. Therefore, there are numerous references related to the calculations of spectral and integral characteristics of blackbody radiation in the finite frequency range of spectrum, including useful tables of Planck functions and their various ratios, as well as a number of important integrals [4-12]. In the latter case, the integrals are calculated numerically or presented in terms of infinite summations. Some approximations are also used to replace the Planck function by simple expressions, which can then be easily integrated [13, 14]. Using polylogarithm formulation allow us a number of integrals of the Planck function within a finite range of frequencies to calculate analytically [4, 15-18].

The main objective of previous research has focused only in obtaining large amounts of data on the radiative properties of blackbody radiation in finite spectral ranges of the spectrum. Nevertheless, currently, there is no data in the literature on the analytical expressions of the thermodynamic functions, such as the Helmholtz free energy density, internal energy density, enthalpy density, entropy density, heat capacity at constant volume and pressure in the different spectral bands of the Planck spectrum.

Thus, it is desirable to obtain the analytical expressions for these functions and, as a consequence, their values for different frequency ranges can be presented in tabular forms.

The present work is devoted to the construction of the exact analytic expressions for the thermodynamic functions of blackbody radiation, as well as a number density of photons in the finite frequency ranges. Direct way to do this is to use the polylogarithm functions. Using these functions the polylogarithmic representation for the Wien displacement law, Helmholtz free energy density, enthalpy density, internal energy density, entropy density, heat capacity at constant volume, pressure, and a number density of photons are obtained. To obtain their values, as well as the ones of radiative functions, such as Stefan-Boltzmann law and total energy density at various temperatures, the values of $B(x_1, x_2)$, $C(x_1, x_2)$, $D(x_1, x_2)$, and $E(x_1, x_2)$ functions for various ranges of frequencies are calculated and presented in Table 1.

As an example, the Stefan-Boltzmann law, the total energy density, and thermodynamic functions are calculated using experimental data for the monopole spectrum of CMB radiation measured by the *COBE* FIRAS instrument in the 60 - 600 GHz frequency range at the temperature $T = 2.725$ K.

The expressions obtained for the radiative and thermodynamic functions can be easily presented in wavelength and wavenumber domains.

## 2. Planck function and Wien's displacement law in a finite frequency range

It is well knows that the spectrum of blackbody radiation is described by the Planck formula [19]

$$I^P(\nu, T) = \frac{8\pi h}{c^3} \frac{\nu^3}{e^{\frac{h\nu}{k_B T}} - 1} . \tag{1}$$

Here $k_B$ is the Boltzmann constant, $T$ is the temperature, $c$ is the light velocity, and $h$ is the Planck constant.

Using expression for the polylogarithms of zero [20]

$$\text{Li}_0(x) = \frac{x}{1-x} , \qquad |x| < 1 , \tag{2}$$

Eq. 1 can be rewritten in the polylogarithmic representation in the form

$$I^P(\nu, T) = \frac{8\pi h \nu^3}{c^3} \text{Li}_0(e^{-\frac{h\nu}{k_B T}}) \tag{3}$$

It is well-known [19] that for obtaining the Wien displacement law, we have to solve the transcendental equation ($\frac{\partial^P I(\nu,T)}{\partial \nu} = 0$) numerically. There is the alternative method so-called *median* one to find the wavelength where a maximum of Planck function occurs [21]. However, to determine the maximum of the Planck function, we have to solve the resulting equation numerically.

In [16], another alternative method for determining of the peak of the Planck function using the ratio of the first and zero moments of the distribution function, was described. This method allows to obtain the analytical expression for the Wien displacement law for the fractional centroid peak.

A method of moments to determine the "generalized" Wien displacement law of real bodies was proposed in [22]. The true temperature is determined by the position of the maximum of the Planck function. The performance of this method was demonstrated on the thermal radiation of tungsten, molybdenum, luminous flames, and zirconium, hafnium, and titanium carbides, and $ZrB_2$-SiC-based ultra-high temperature ceramics [22-27].

Let us construct the "generalized" Wien displacement law in the polylogarithmic representation. According to [22], Wien's displacement law of blackbody radiation can be represented as

$$X_{max} = \frac{h\nu_{max}}{k_B T} = \beta(1-\gamma), \qquad (4)$$

where

$$\beta = \frac{h}{k_B T} \cdot \frac{S_1}{S_0}, \quad \gamma = \frac{S_0 \bar{S}_3}{2 S_1 \bar{S}_2}. \qquad (5)$$

Here $S_n(\nu_1, \nu_2, T)$ and $\bar{S}_n(\nu_1, \nu_2, T)$ are determined as follows

$$S_n(T) = \frac{8\pi h}{c^3} \int_{\nu_1}^{\nu_2} \frac{\nu^{3+n}}{e^{\frac{h\nu}{k_B T}} - 1} d\nu \qquad (6)$$

$$\bar{S}_n(T) = \frac{8\pi h}{c^3} \int_{\nu_1}^{\nu_2} (\nu - \bar{\nu})^n \frac{\nu^3}{e^{\frac{h\nu}{k_B T}} - 1} d\nu. \qquad (7)$$

Eq. 6 and Eq. 7 are the initial and central moments of the distribution function $I^P(\nu,T)$, n = 0, 1, 2, 3 …k is the order of the moments; and $\bar{\nu} = \dfrac{S_1}{S_0}$. As seen in Eq. (7), the central moments of the distribution function are defined through a combination of initial moments.

Using the polylogarithm functions [20], Eq. 6 and Eq. 7 can be rewritten as follows:

$$S_n(T) = \frac{8\pi h}{c^3} \int_{\nu_1}^{\nu_2} \nu^n \mathrm{Li}_0(e^{-\frac{h\nu}{k_B T}}) d\nu \qquad (8)$$

$$\bar{S}_n(T) = \frac{8\pi h}{c^3} \int_{\nu_1}^{\nu_2} (\nu-\bar{\nu})^n \mathrm{Li}_0(e^{-\frac{h\nu}{k_B T}}) d\nu. \qquad (9)$$

By integrating Eq. 8 over a given spectral band, we obtain

$$S_n(x_1, x_2) = \frac{8\pi (k_B T)^{4+n}}{c^3 h^{3+n}} (3+n)! A_n(x_1, x_2), \qquad (10)$$

where

$$A_n(x_1, x_2) = [P_{3+n}(x_1) - P_{3+n}(x_2)]. \qquad (11)$$

Here $x = \dfrac{h\nu}{k_B T}$ and $P_3(x)$ is defined as

$$P_{3+n}(x) = \sum_{s=0}^{3+n} \frac{(x)^s}{s!} \mathrm{Li}_{4+n-s}(e^{-x}), \qquad (12)$$

where

$$\mathrm{Li}_{4+n-s}(e^{-x}) = \sum_{k=1}^{\infty} \frac{e^{-kx}}{k^{4+n-s}}, \quad |e^{-kx}| < 1 \qquad (13)$$

is the polylogarithm functions [20].

Substituting the expressions for the initial and central moments Eqs. 9 and 10 in Eq. 4, Wien's displacement law has the structure

$$X_{\max} = \frac{4(P_4(x_1) - P_4(x_2))}{P_3(x_1) - P_3(x_2)} \left[1 - \gamma(S_0(x_1, x_2), S_1(x_1, x_2), \bar{S}_2(x_1, x_2), \bar{S}_3(x_1, x_2))\right] \qquad (14)$$

Such representation of the Wien displacement law allow us to calculate it at different temperatures analytically.

Now, let's obtain Wien's displacement law for the entire spectrum of blackbody radiation ($0 \leq \nu \leq \infty$).

In this case, since $P_{3+n}(\infty) = 0$ and $P_{3+n}(0) = \text{Li}_{4+n}(1)$, Eq. 10 for the $n^{th}$ initial moment of the Planck function takes the form

$$S_n(0, \infty) = \frac{8\pi (k_B T)^{n+4}}{c^3 h^{3+n}} (3+n)! \, \text{Li}_{4+n}(1) \, . \tag{15}$$

Here $\text{Li}_{4+n}(1) = \varsigma(n+5)$, $\varsigma(n+5)$ is the Riemann zeta function.

In accordance with Eq. 9, the central moments defined as follows

$$\bar{S}_0 = S_0 \tag{16}$$

$$\bar{S}_1 = 0 \tag{17}$$

$$\bar{S}_2 = S_2 - \frac{S_1^2}{S_0} \tag{18}$$

$$\bar{S}_3 = S_3 - 3\frac{S_1 S_2}{S_0} + 2\frac{S_1^3}{S_0^2} \, . \tag{19}$$

Substituting the analytical expressions for initial and central moments Eqs. 15-19 in Eq. 14, Wien's displacement law has the structure

$$X_{\max} = \beta'\left[1 - \gamma'(S_0(0,\infty), S_1(0,\infty), \bar{S}_2(0,\infty), \bar{S}_3(0,\infty))\right] , \tag{20}$$

where

$$\beta' = \frac{4(P_4(0) - P_4(\infty))}{P_3(0) - P_3(\infty)} = \frac{4\text{Li}_5(1)}{\text{Li}_4(1)} = 3.83222 \tag{21}$$

and

$$\gamma' = 0.26103 \, . \tag{22}$$

Finally, Wien's displacement law has the following expression:

$$X_{max} = 2.8318 \ . \tag{23}$$

As we can see, Eq. 23 negligibly different from the well-known Wien's displacement law ($X_{max} = 2.8214$) [19]. The uncertainty in this case does not exceed 1%.

In conclusion, we note the following. If we will determine the position of the maximum of the Planch function using the ratio of the first and zero moments

$$X_{max} = \beta' = \frac{h}{k_B T} \frac{S_1}{S_0} \tag{24}$$

then, as can be seen from Eq. 20, the uncertainty is more than 20%. Therefore, to obtain a more accurate value of Wien's displacement law, Eq. 4, the higher-orders moments should be taken into account.

### 3. Number density of photons in the finite range of frequencies

The number density of photons of blackbody radiation with a photon energy from $h\nu_1$ to $h\nu_2$, is defined as follows [19].

$$n = \frac{8\pi}{c^3} \int_{\nu_1}^{\nu_2} \frac{\nu^2}{e^x - 1} d\nu \ . \tag{25}$$

In accordance with Eq. 25, the polylogarithmic representation of the number density of photons has the following form:

$$n = \frac{8\pi}{c^3} \int_{\nu_1}^{\nu_2} \nu^2 \text{Li}_0(e^{-\frac{h\nu}{k_B T}}) d\nu = \frac{16\pi k_B^3}{c^3 h^3} T^3 B(x_1, x_2) \tag{26}$$

$$B(x_1, x_2) = [P_2(x_1) - P_2(x_2)] \tag{27}$$

The values of $B(x_1, x_2)$ function for the different variables $x_1$ and $x_2$ are presented in Table 1.

In the semi-infinite range of frequencies, Eq. 26 takes the form

$$n = \frac{8\pi}{c^3} \int_0^\infty \nu^2 \text{Li}_0(e^{-\frac{h\nu}{k_B T}}) d\nu = \frac{16\pi k_B^3}{c^3 h^3} T^3 [P_2(0) - P_2(\infty)]. \tag{28}$$

Since $P_2(0) = \text{Li}_3(1) = \zeta(3) \approx 1.202$ and $P_2(\infty) = 0$, Eq. 28 transits to the well-known expression [19]

$$n = \frac{8\pi}{c^3}\int_0^\infty v^2 \text{Li}_0(e^{-\frac{hv}{k_B T}})dv \approx 19.232 \frac{\pi k_B^3}{c^3 h^3}T^3 \ . \tag{29}$$

## 4. The polylogarithmic representation of the Stefan-Boltzmann law

According to Eq. 3, the polylogarithmic representation of the total energy density in the finite frequency range from $v_1$ to $v_2$ has the form

$$I(v_1, v_2, T) = \frac{8\pi h}{c^3}\int_{v_1}^{v_2} v^3 \text{Li}_0(e^{-\frac{hv}{k_B T}})dv \tag{30}$$

To obtain a solution for Planck integral, Eq. 30, we have to compute the integral in the finite frequency range from $v_1$ to $v_2$, which give the following solution after integration:

$$I(x_1, x_2, T) = S_0(x_1, x_2, T) = \frac{8\pi h}{c^3}\int_{v_1}^{v_2} v^3 \text{Li}_0(e^{-\frac{hv}{k_B T}})dv = \frac{48\pi(k_B T)^4}{c^3 h^3} C(x_1, x_2) \ , \tag{31}$$

where

$$C(x_1, x_2) = [P_3(x_1) - P_3(x_2)]. \tag{32}$$

If the spectral range is extended to the entire spectrum $0 \leq v \leq \infty$, Eq. 31 can be rewritten as follows

$$I(0, \infty, T) = \frac{48\pi k_B^4}{c^3 h^3}T^4(P_3(0) - P_3(\infty)) = aT^4, \tag{33}$$

where $a$ is the radiation density constant.

Since $P_3(0) = \text{Li}_4(1) = \zeta(4) = \frac{\pi^4}{90}$ and $P_3(\infty) = 0$, the radiation density constant has a well-known value

$$a = \frac{24\pi^5 k_B^4}{45 c^3 h^3} = 7.5657 \times 10^{-16} \frac{\text{J}}{\text{m}^3 \text{K}^4} \text{ [19]}.$$

Using relationship between the total energy density and the total radiation power per unit area $I^{SB} = \frac{c}{4}I$, the Stefan-Boltzmann law in the finite frequency range has the form

$$I^{SB}(x_1, x_2, T) = \frac{2\pi h}{c^2}\int_{v_1}^{v_2} v^3 \text{Li}_0(e^{-\frac{hv}{k_B T}})dv = \frac{12\pi(k_B T)^4}{c^2 h^3} C(x_1, x_2) \ . \tag{34}$$

$C(x_1, x_2)$ is determined by Eq. 32.

For the entire spectrum of Planck function Eq. 34 transits to a well-knows expression

$$I^{SB}(0,\infty,T) = \frac{12\pi(k_B T)^4}{c^2 h^3}[P_3(0) - P_3(\infty)] = \frac{2\pi^5 k_B^4}{15 c^2 h^3} T^4 = \sigma T^4, \tag{35}$$

where $\sigma$ is the Stefan-Boltzmann constant. The value of latter is [19]:

$$\sigma = \frac{2\pi^5 k_B^4}{15 c^2 h^3} = 5.67037 \times 10^{-8} \frac{W}{m^2 K^4}. \tag{36}$$

The calculated values of $C(x_1, x_2)$ in different frequency bands are shown in Table 1.

## 5. Thermodynamic functions in finite range of frequencies

The thermodynamic functions of the blackbody radiation in the finite frequency range are determined by the following expressions [19]:

1) Helmholtz free energy density $f = \frac{F}{V}$:

$$f(v_1, v_2, T) = \frac{8\pi k_B T}{c^3} \int_{v_1}^{v_2} v^2 \ln\left(1 - e^{\frac{hv}{k_b T}}\right) dv \tag{37}$$

2) Entropy density $s = \frac{S}{V}$:

$$s = -\frac{\partial f}{\partial T} \tag{38}$$

3) Heat capacity at constant volume per unit volume $c_V = \frac{C_V}{V}$:

$$c_V = \left(\frac{\partial I(v_1, v_2, T)}{\partial T}\right)_V \tag{39}$$

4) Pressure of photons $P$:

$$P = -f. \tag{40}$$

As can be seen, in order to build the thermodynamic functions of blackbody radiation Eqs. 37-40, the knowledge of the total energy density and Helmholtz free energy density are necessary.

After changing the variable $x = \dfrac{h\nu}{k_B T}$ in Eq. 37 and the sequence of integration and differentiation, the Eqs. 37 - 40 can be rewritten as follows polylogarithmic presentation:

(1) Helmholtz free energy density $f$:

$$f(x_1, x_2, T) = -\frac{16\pi k_B^4}{c^3 h^3} T^4 C(x_1, x_2) , \qquad (41)$$

where

$$C(x_1, x_2) = P_3(x_1) - P_3(x_2) \qquad (42)$$

(2) Entropy density $s$:

$$s(x_1, x_2) = \frac{64\pi k_B^4}{c^3 h^3} T^3 D(x_1, x_2), \qquad (43)$$

where

$$D(x_1, x_2) = \left[ P_3(x_1) - P_3(x_2) + \frac{1}{24}(x_1^4 \, \mathrm{Li}_0(e^{-x_1}) - x_2^4 \, \mathrm{Li}_0(e^{-x_2})) \right] \qquad (44)$$

(3) Heat capacity at constant volume per unit volume, $c_V$

$$c_V(x_1, x_2) = \frac{192\pi k_B^4}{c^3 h^3} T^3 E(x_1, x_2), \qquad (45)$$

where

$$E(x_1, x_2) = \left[ P_3(x_1) - P_3(x_2) + \frac{1}{24}(x_1^4 \, \mathrm{Li}_0(e^{-x_1}) - x_2^4 \, \mathrm{Li}_0(e^{-x_2})) + \frac{1}{72}(x_1^5 \, \mathrm{Li}_{-1}(e^{-x_1}) - x_2^5 \, \mathrm{Li}_{-1}(e^{-x_2})) \right] \qquad (46)$$

(4) Pressure $p$:

$$p(x_1, x_2) = \frac{16\pi k_B^4}{c^3 h^3} T^4 C(x_1, x_2) , \qquad (47)$$

where

$$C(x_1, x_2) = P_3(x_1) - P_3(x_2). \qquad (48)$$

It is not difficult to show that Eqs. 41, 43, 45, 47 in the semi-infinite range of frequencies are converted to well-known expressions for the thermodynamic functions [19]. For example, using the values of $P_3(0) = \text{Li}_4(1) = \xi(4) = \dfrac{\pi^4}{90}$ and $P_3(\infty) = 0$, a well-known expression for Helmholtz free energy density has the form $f(0,\infty,T) = -\dfrac{1}{3}bT^4$, where $b = -\dfrac{8\pi^5 k_B^4}{15h^3c^3}$ [19].

By definition, $f = u - Ts$ (where $u$ is the internal energy density), we have the following expression for $u = f + Ts$. By combining Eqs. 41 and 43, we obtain:

$$u(x_1, x_2, T) = \dfrac{64\pi k_B^4}{c^3 h^3} T^4 D(x_1, x_2) \left[1 - \dfrac{C(x_1, x_2)}{4D(x_1, x_2)}\right] \qquad (49)$$

The enthalpy density $h$ follows from its definition, $h = u + p$, giving the following form

$$h(x_1, x_2, T) = \dfrac{64\pi k_B^4}{c^3 h^3} T^4 D(x_1, x_2) \qquad (50)$$

The Gibbs free energy density $g$, by definition, is $g = h - Ts$, thus in accordance with Eqs. 43 and 50 is zero.

$$g(x_1, x_2, T) = 0 \qquad (51)$$

By definition [19], the chemical potential density $\mu = \left(\dfrac{\partial f}{\partial n}\right)_{T,V}$, as clearly seen from Eq. 41, is equal to zero too.

$$\mu(x_1, x_2, T) = 0 \qquad (52)$$

The calculated values of $B(x_1, x_2)$, $C(x_1, x_2)$, $D(x_1, x_2)$, $E(x_1, x_2)$ functions for different frequency ranges are presented in Table 1. The values of $x_1$ and $x_2$ have been chosen only for eliminating the length on the present manuscript.

In conclusion, let us consider an example related to the calculation of radiative and thermodynamic functions of the Cosmic Microwave Background (CMB) radiation. It is well-known that the theoretical prediction that the CMB spectrum is so close to a blackbody was confirmed by the *COBE* FIRAS observation [28]. Currently, only this data measured by the *COBE* FIRAS instrument within the finite range of frequencies, ranging from 60 to 600 GHz is reliable for the calculation of the radiative and thermodynamic functions of the CMB radiation. Now let us apply the expressions obtained for calculating the radiative and thermodynamic functions of the CMB radiation for the monopole spectrum (blackbody spectrum) at the temperature $T = 2.725$ K [29]. Using the measured data obtained by the

*COBE* FIRAS instrument in the finite range of frequencies, the radiative and thermodynamic properties of the CMB radiation are calculated and presented in Table 2.

In conclusion, it is important to note that the exact analytical expressions obtained above for the radiative and thermodynamic functions of blackbody radiation in the finite range of frequencies can be easily presented in the wavenumber ($\lambda$) or wavelength ($\tilde{v}$) domains. In this case, we should use the following relationships [30]:

$$v = \frac{c}{\lambda} \ , \quad v = c\tilde{v} \tag{53}$$

$$dv = -\frac{c}{\lambda^2} d\lambda \ , \quad dv = cd\tilde{v} \tag{54}$$

$$\int_{v_1}^{v_2} I^P(v,T)dv = \int_{\lambda_2}^{\lambda_1} I^P(\lambda,T)d\lambda = \int_{\tilde{v}_1}^{\tilde{v}_2} I^P(\tilde{v},T)d\tilde{v} \tag{55}$$

## 6. Conclusions

In this paper, using the polylogarithm functions the analytic expressions for the radiative and thermodynamic functions of blackbody radiation, such as the Wien displacement law, Stefan-Boltzmann law, total energy density, number density of photons, Helmholtz free energy density, enthalpy density, internal energy density, entropy density, heat capacity at constant volume, pressure, enthalpy density, and internal energy density in the finite range of frequencies are obtained. In the case of the entire range of the Planck spectrum, these functions are transited to the well-known expressions [19].

For obtaining the radiative and thermodynamic properties of blackbody radiation for various spectral bands of the Planck spectrum at different temperatures, the values of $B(x_1,x_2)$, $C(x_1,x_2)$, $D(x_1,x_2)$, and $E(x_1,x_2)$ functions are calculated and presented in Table 1.

Utilizing the experimental data for the monopole spectrum measured by the *COBE* FIRAS instrument in the finite range of frequencies $60\text{GHz} \leq v \leq 600\text{GHz}$ at the temperature $T = 2.725$ K, the values of the radiative and thermodynamic functions are calculated and presented in Table 2.

In conclusion it is important to note that there are several classes of materials and objects for which the polylogarithmic representation of the radiative and thermodynamic properties of thermal radiation takes place. As examples, we can provide the following real bodies and objects: a) luminous flames [31]; b) cobalt, iron, and nickel at their melting points [32]; c) Milky Way and other galaxies [33]; and others. As a result, the analytical expressions of the radiative and thermodynamic functions for real bodies and objects can be constructed and their values for different frequency ranges at different temperatures can be tabulated.

These and other topics will be points of discussion in subsequent publications.

**Acknowledgments**

The authors cordially thank Professor N.P. Malomuzh for fruitful discussions.

| $x_1$ | $x_2$ | $B(x_1, x_2)$ | $C(x_1, x_2)$ | $D(x_1, x_2)$ | $E(x_1, x_2)$ |
|---|---|---|---|---|---|
| 0.1 | 1.1 | 0.204215951 | 0.137199356 | 0.070149439 | 0.017400406 |
| - | 2.1 | 0.521708648 | 0.418448644 | 0.26752365 | 0.10417675 |
| - | 3.1 | 0.791586525 | 0.677301531 | 0.505941818 | 0.267334359 |
| - | 4.1 | 0.974339954 | 0.857323639 | 0.713645667 | 0.46737761 |
| - | 5.1 | 1.082865989 | 0.965252119 | 0.864144273 | 0.657569929 |
| - | 6.1 | 1.141931106 | 1.024197217 | 0.960783723 | 0.810001354 |
| - | 7.1 | 1.172149197 | 1.054392648 | 1.017757165 | 0.918149049 |
| - | 8.1 | 1.186917678 | 1.069157043 | 1.049283861 | 0.988239368 |
| - | 9.1 | 1.193887559 | 1.076126214 | 1.065901133 | 1.030622668 |
| - | 10.1 | 1.197087902 | 1.079326436 | 1.07432981 | 1.054891675 |
| 1.1 | 2.1 | 0.317492697 | 0.281249288 | 0.197374211 | 0.086776344 |
| - | 3.1 | 0.587370574 | 0.540102175 | 0.435792379 | 0.249933953 |
| - | 4.1 | 0.770124004 | 0.720124284 | 0.643496227 | 0.449977204 |
| - | 5.1 | 0.878650038 | 0.828052763 | 0.793994834 | 0.640169523 |
| - | 6.1 | 0.937715155 | 0.886997861 | 0.890634283 | 0.792600948 |
| - | 7.1 | 0.967933246 | 0.917193292 | 0.947607726 | 0.900748643 |
| - | 8.1 | 0.982701727 | 0.931957688 | 0.979134421 | 0.970838962 |
| - | 9.1 | 0.989671609 | 0.938926859 | 0.995751694 | 1.013222262 |
| - | 10.1 | 0.992871951 | 0.942127081 | 1.004180371 | 1.037491269 |
| 2.1 | 3.1 | 0.269877877 | 0.258852887 | 0.238418168 | 0.163157609 |
| - | 4.1 | 0.452631307 | 0.438874996 | 0.446122017 | 0.36320086 |
| - | 5.1 | 0.561157341 | 0.546803475 | 0.596620623 | 0.553393179 |
| - | 6.1 | 0.620222458 | 0.605748573 | 0.693260073 | 0.705824604 |
| - | 7.1 | 0.650440549 | 0.635944004 | 0.750233515 | 0.813972299 |
| - | 8.1 | 0.66520903 | 0.6507084 | 0.78176021 | 0.884062618 |
| - | 9.1 | 0.672178912 | 0.657677571 | 0.798377483 | 0.926445918 |
| - | 10.1 | 0.675379254 | 0.660877792 | 0.80680616 | 0.950714925 |
| 3.1 | 4.1 | 0.18275343 | 0.180022109 | 0.207703849 | 0.200043251 |
| - | 5.1 | 0.291279464 | 0.287950588 | 0.358202455 | 0.39023557 |
| - | 6.1 | 0.350344581 | 0.346895686 | 0.454841904 | 0.542666995 |
| - | 7.1 | 0.380562672 | 0.377091117 | 0.511815347 | 0.65081469 |
| - | 8.1 | 0.395331153 | 0.391855513 | 0.543342042 | 0.720905009 |
| - | 9.1 | 0.402301035 | 0.398824683 | 0.559959315 | 0.763288309 |
| - | 10.1 | 0.405501377 | 0.402024905 | 0.568387992 | 0.787557316 |
| 4.1 | 5.1 | 0.108526035 | 0.10792848 | 0.150498606 | 0.190192319 |
| - | 6.1 | 0.167591152 | 0.166873578 | 0.247138056 | 0.342623743 |
| - | 7.1 | 0.197809243 | 0.197069009 | 0.304111499 | 0.450771439 |
| - | 8.1 | 0.212577723 | 0.211833404 | 0.335638194 | 0.520861757 |
| - | 9.1 | 0.219547605 | 0.218802575 | 0.352255467 | 0.563245057 |
| - | 10.1 | 0.222747947 | 0.222002797 | 0.360684143 | 0.587514065 |
| 5.1 | 6.1 | 0.059065117 | 0.058945098 | 0.09663945 | 0.152431425 |
| - | 7.1 | 0.089283208 | 0.089140529 | 0.153612892 | 0.26057912 |
| - | 8.1 | 0.104051688 | 0.103904924 | 0.185139587 | 0.330669438 |
| - | 9.1 | 0.11102157 | 0.110874095 | 0.20175686 | 0.373052738 |
| - | 10.1 | 0.114221912 | 0.114074317 | 0.210185537 | 0.397321746 |
| 6.1 | 7.1 | 0.030218091 | 0.030195431 | 0.056973443 | 0.108147695 |
| - | 8.1 | 0.044986572 | 0.044959826 | 0.088500138 | 0.178238014 |
| - | 9.1 | 0.051956453 | 0.051928997 | 0.105117411 | 0.220621314 |
| - | 10.1 | 0.055156796 | 0.055129219 | 0.113546087 | 0.244890321 |
| 7.1 | 8.1 | 0.014768481 | 0.014764395 | 0.031526695 | 0.070090319 |
| - | 9.1 | 0.021738362 | 0.021733566 | 0.048143968 | 0.112473619 |
| - | 10.1 | 0.024938705 | 0.024933788 | 0.056572645 | 0.136742626 |
| 8.1 | 9.1 | 0.006969882 | 0.006969171 | 0.016617273 | 0.0423833 |
| - | 10.1 | 0.010170224 | 0.010169393 | 0.02504595 | 0.066652307 |
| 9.1 | 10.1 | 0.003200342 | 0.003200222 | 0.008428677 | 0.024269008 |

**Table 1** Calculated values of the following functions: $B(x_1, x_2)$, $C(x_1, x_2)$, $D(x_1, x_2)$, $E(x_1, x_2)$.

| Quantity | Monopole (blackbody) spectrum $60\,\text{GHz} \leq \nu \leq 600\,\text{GHz}$ |
|---|---|
| $I(\nu_1, \nu_2, T)$ $[\text{J m}^{-3}]$ | $3.984 \times 10^{-14}$ |
| $I^{SB}(\nu_1, \nu_2, T)$ $[\text{W m}^{-2}]$ | $2.984 \times 10^{-6}$ |
| $f$ $[\text{J m}^{-3}]$ | $-1.226 \times 10^{-14}$ |
| $u$ $[\text{J m}^{-3}]$ | $3.981 \times 10^{-14}$ |
| $h$ $[\text{J m}^{-3}]$ | $4.881 \times 10^{-14}$ |
| $s$ $[\text{J m}^{-3}\,\text{K}^{-1}]$ | $1.911 \times 10^{-14}$ |
| $p$ $[\text{J m}^{-3}]$ | $1.226 \times 10^{-14}$ |
| $c_V$ $[\text{J m}^{-3}\,\text{K}^{-1}]$ | $5.924 \times 10^{-14}$ |
| $n$ $[\text{m}^{-3}]$ | $3.440 \times 10^{8}$ |

**Table 2** Calculated values of the radiative and thermodynamic state functions for the monopole spectrum (blackbody spectrum) in the 60 – 600 GHz frequency interval and semi-infinite interval at $T = 2.725$ K.